\documentclass[aps,pre,twocolumn,nofootinbib,groupedaddress,showpacs,preprintnumbers,notitlepage]{revtex4-1}
\usepackage{amsmath,amsfonts,amssymb}
\usepackage{bm,color}
\usepackage[dvipsnames]{xcolor}
\usepackage{graphicx}
\usepackage{comment}
\usepackage{placeins}

\let\revappendix\appendix

\usepackage[capitalize]{cleveref}

\newcommand{\globaldensity}{\ensuremath{\rho_0}}
\newcommand{\globaldensitycrit}{\ensuremath{\rho_{0,c}}}
\renewcommand{\vec}[1]{\ensuremath{\mathbf{#1}}}
\renewcommand{\matrix}[1]{\ensuremath{\mathcal{#1}}}

\begin{document}

\title{Interpretable machine learning for inferring the phase boundaries in a nonequilibrium system}

\author{C.~Casert}
\email{corneel.casert@ugent.be}
\author{T.~Vieijra}
\author{J.~Nys}
\author{J.~Ryckebusch}
\affiliation{Department of Physics and Astronomy, Ghent University, 9000 Ghent, Belgium}

\begin{abstract}
Still under debate is the question of whether machine learning is capable of going beyond black-box modeling for complex physical systems.
 We investigate the generalizing and interpretability properties of learning algorithms. To this end, we use supervised and unsupervised learning to infer the phase boundaries of the active Ising model, starting from an ensemble of configurations of the system. We illustrate that unsupervised learning techniques are powerful at identifying the phase boundaries in the control parameter space, even in situations of phase coexistence. It is demonstrated that supervised learning with neural networks is capable of learning the characteristics of the phase diagram, such that the knowledge obtained at a limited set of control variables can be used to determine the phase boundaries across the phase diagram. In this way, we show that properly designed supervised learning provides predictive power to regions in the phase diagram that are not included in the training phase of the algorithm. We stress the importance of introducing interpretability methods in order to perform a physically relevant classification of the phases with deep learning. 
\end{abstract}

\maketitle

\section{Introduction} Machine learning has recently shown its great potential for addressing nontrivial problems in statistical and many-body physics. Successful applications include the detection of phase transitions in spin systems~\cite{Carrasquilla2017, vanNieuwenburg2017, PhysRevLett.120.176401,PhysRevLett.120.066401, PhysRevX.7.031038, PhysRevB.97.045207, PhysRevB.96.184410, PhysRevB.97.174435, PhysRevLett.120.257204, PhysRevB.98.060301, PhysRevB.94.195105, PhysRevE.95.062122,PhysRevE.96.022140,PhysRevE.97.013306, PhysRevE.97.032119, PhysRevE.98.022138}, mapping the ground-state wave function of quantum many-body systems and performing quantum state tomography~\cite{Carleo602,Torlai2018, PhysRevLett.120.240503}, exploiting the apparent similarities between neural networks and the theory of the renormalization group~\cite{mehta2014exact,Koch-Janusz2018, huang2017neural, PhysRevX.7.021021, PhysRevB.97.085104, PhysRevE.97.053304, PhysRevLett.121.260601}, and the acceleration of Monte Carlo simulations~\cite{PhysRevB.95.041101, PhysRevB.95.035105, PhysRevB.95.241104, PhysRevE.96.051301}.\\

Due to its expressive power, deep learning has proven to be a powerful tool to identify phase boundaries. Yet, interpretability---i.e., can we understand on what the machine learning algorithm bases its decision?---remains an issue. Indeed, it often remains unclear how to transfer the features identified by a neural network to comprehensible physical properties. Thereby it cannot be excluded that the neural network does not even learn physically relevant properties altogether. Thus far, interpretable machine-learning methods for physical systems have often drawn on the use of more transparent (albeit less expressive) learning methods, such as support vector machines~\cite{PhysRevB.96.205146,greitemann2018probing}.
 Although promising, relatively few studies have been devoted to deepening our understanding of the properties of a physical system with the aid of deep learning~\cite{PhysRevLett.120.066401, PhysRevB.96.184410}. Gaining a more general insight into whether a deep neural network's classification can be built on nontrivial physical features would hence be a major step forward in the development of an interpretable deep learning methodology for selected physics applications.\\
 
In this work, we sketch a possible road map for such an interpretable learning methodology that is capable of inferring the high-level features of a system in the control parameter space by merely starting from an ensemble of system configurations. We propose a two-step procedure: ($i$) first, we apply unsupervised learning to identify the phase boundaries in a slice of the phase diagram, ($ii$) subsequently, we use supervised methods to extract the relevant features of the phases labeled in step ($i$). Thereby, we show how to select specific models that can learn characteristic properties of the physical system, and we use these to complete the phase diagram. We show that the aptitude of the neural network to classify phases in a physically relevant fashion can be considerably enhanced by introducing interpretability tools that provide an improved comprehension of the internal representation of the networks.\\

As a prototypical example, we apply our methodology to configurations of the active Ising model (AIM)~\cite{PhysRevLett.111.078101,PhysRevLett.114.068101,PhysRevE.92.042119}, a nonequilibrium spin system with a nontrivial phase diagram. The AIM describes the generic features of collective motion emerging from local interactions in a lattice gas. Collective motion has played a preeminent role in the study of active matter, and the flocking transition has attracted widespread attention due to its universal properties~\cite{PhysRevLett.75.1226,PhysRevE.77.046113,VICSEK201271}. The nature of this phase transition has  been established to be comparable to a liquid-gas transition.  In the two-dimensional AIM, particles with spin projections $s = \pm 1$ undergo biased diffusion along the $x$-axis, and diffuse freely along the $y$-axis. Particles hop to the left (right) at a rate $D(1\mp\epsilon s)$, where $D$ is a diffusion coefficient and $\epsilon$ is a measure for the self-propulsion. Hopping along the $y$-axis is symmetric at a rate $D$. The number $n_{\pm,i} $ of $s=\pm 1$ spins on a lattice site $i$ determines the local density $\rho _i = n_{+,i} + n_{-,i}$ and the local magnetization $m_i = n_{+,i} -n_{-,i}$.
Particles on a particular lattice site $i$ tend to align their spin through a ferromagnetic interaction: a spin flip occurs at a rate $\exp(-s\beta m_i/\rho_i)$, with $\beta = 1/T$ the inverse temperature. The global density $\globaldensity = \sum_i \rho_i/L^2$ is fixed. The control parameter space of the dynamic system under investigation is defined by the variables $(\globaldensity, T, D, \epsilon)$. At a fixed $\epsilon>0$ and $D>0$, the phase diagram in the $(\globaldensity, T)$-plane has three distinct regions~\cite{PhysRevLett.111.078101,PhysRevLett.114.068101,PhysRevE.92.042119}. For low $\globaldensity$ and high $T$, collective motion is absent and the system behaves like a homogeneous gas (phase `G') with mean magnetization per spin $\overline{m} \approx 0$, where  \begin{equation}\label{eq:m}\overline{m} = \frac{1}{\globaldensity L^2} \left|\sum_{i=1}^{L^2} m_i\right|. \end{equation} At high $\globaldensity$ and low $T$, the system acts like a homogeneous polar liquid (phase `L'). For intermediate values of $\globaldensity$ and $T$, phase separation is observed in the form of an ordered, high-density band moving through a disordered, dilute gas (phase `L+G'). The critical point of this phase transition, where the system can continuously transform between liquid and gas, lies at \mbox{$(\globaldensitycrit = \infty, T_c =1)$} and no supercritical region exists. In the following, we use $D = 1$ and $\epsilon = 0.9$ without any loss of generality, as these variables only affect the precise location of the phase boundaries and not the overall qualitative features of the phase diagram. Our results are obtained for a fixed system size $L = 81$, which is large enough so that all three phases can be observed. The focus of our work is on comparing machine-learning results for phase classification with more traditional approaches for systems at a specific system size $L$. The extraction of the phase diagram in the thermodynamic limit is beyond the scope of our study.\\

The organization of the rest of this work is as follows: in Sec.~II we show that unsupervised learning techniques are capable of clustering  AIM configurations into the various phases, even in the presence of phase coexistence. This is done most accurately with a recently developed technique based on manifold learning.  In Sec.~III, deep learning is used to extrapolate the inferred phase boundaries to other combinations of the control parameters. In this process of completing the phase diagram, we illustrate that the introduction of an interpretability tool is indispensable.

\section{Unsupervised learning} We now explore to what extent unsupervised  machine learning is capable of uncovering the nontrivial phase diagram of the AIM. We use dimensionality reduction methods to identify the relevant subspace of configuration space that characterizes the different phases at varying temperatures $T$ and fixed $\globaldensity=3$. Using unsupervised algorithms such as principal component analysis (PCA)~\cite{PhysRevB.94.195105, PhysRevE.96.022140, PhysRevE.95.062122} and uniform manifold approximation and projection (UMAP)~\cite{mcinnes2018umap}, we illustrate that one can cluster the AIM configurations in groups corresponding to their respective phase. To this end, we introduce the data matrix $\matrix{D}$, containing the local magnetization values of $N$ configurations. $\matrix{D}_{ji}$ represents the magnetization at site $i$ for a configuration $j$. The rows of $\matrix{D}$ correspond to $50$ uncorrelated configurations per temperature $T \in [0.2,1.0]$ with a temperature spacing $\Delta T = 0.01$. Hence, for an $L\times L$ lattice, $\matrix{D}$ is an $N \times L^2$ matrix. The uncorrelated AIM configurations in $\matrix{D}$ are sampled using Markov Chain Monte Carlo. \\ 

\begin{figure}[t!]
\includegraphics{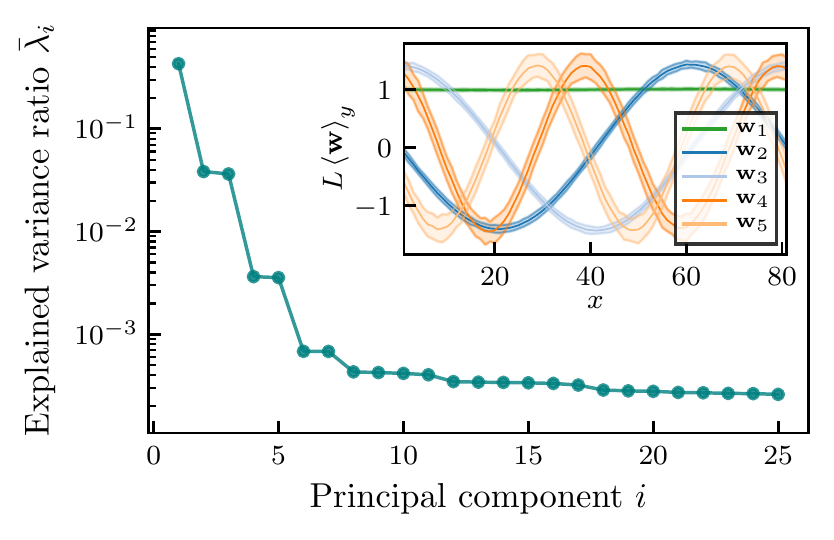}
\caption{The explained variance ratio $\bar{\lambda}_i$ for the first 25 principal components of the data matrix $\mathcal{D}$, for $L=81$ and $\globaldensity = 3$. Inset: The $x$-dependence of the first five principal components, averaged along the $y$-direction. The shaded region corresponds to three standard deviations on this average.}
  \label{fig:expvar}
\end{figure}%
\begin{figure*}
\includegraphics[width=\textwidth]{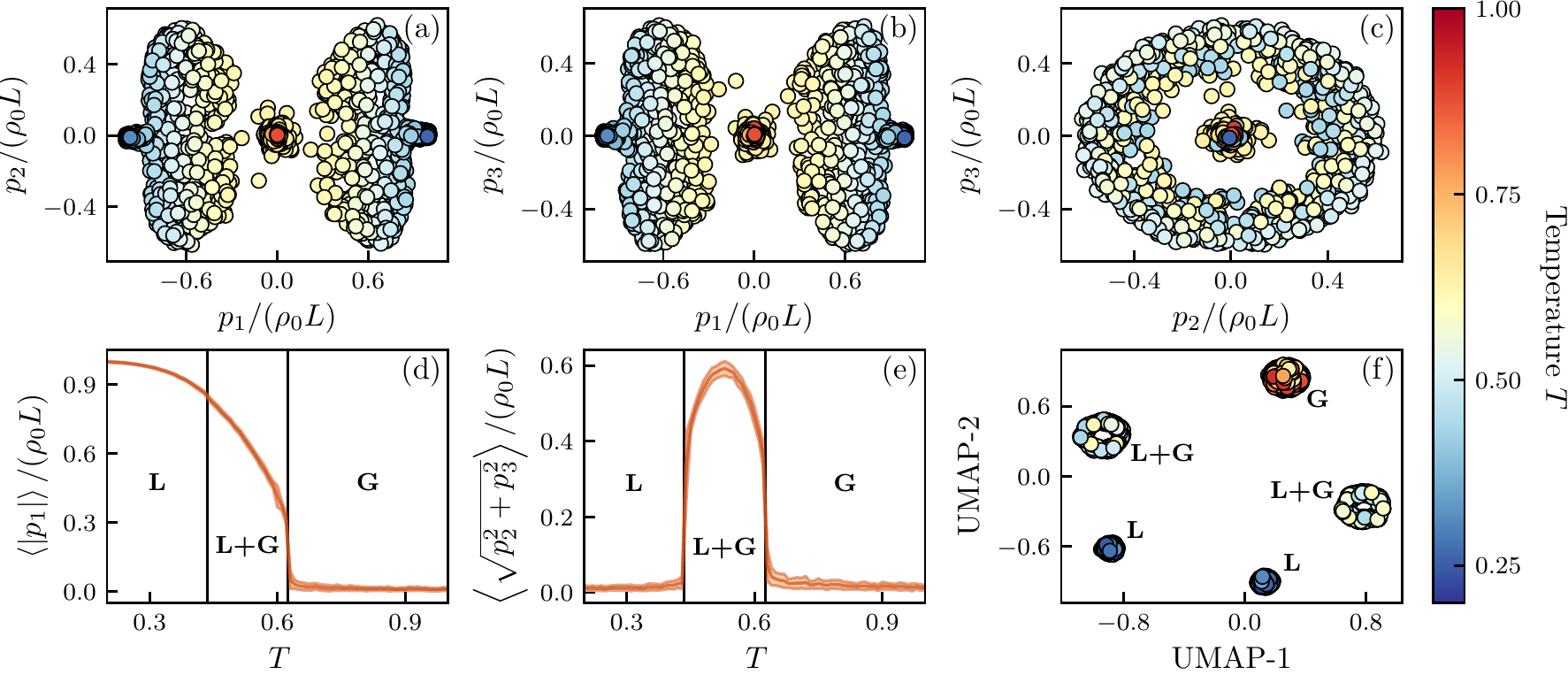}
\caption{Classification of AIM configurations at $\globaldensity = 3$ and various temperatures $0.2 \le T \le 1$ with the unsupervised PCA (panels (a) to (e)) and UMAP (panel (f)) techniques. (a), (b) and (c) Scatter plots with the projection of the 4050 AIM spin configurations on the first three principal components. (d) Fixed-temperature average of $|p_1|$ as a function of $T$. The shaded region corresponds to one standard deviation. The labels indicate the phase boundaries obtained by evaluating the mean magnetization per particle $\overline{m}$ (Eq.~(\ref{eq:m}))  and the liquid fraction $\phi$ (Eq.~(\ref{eq:phi})). (e) Fixed-temperature average of $\sqrt{p_2^2+p_3^2}$ as a function of $T$. (f) Clustering of the AIM configurations with the UMAP algorithm.}
  \label{fig:unsupervised}
\end{figure*}%
PCA identifies the dominant features of a data set  as the orthogonal and linearly uncorrelated variables (principal components) by which its variance can best be explained. The principal components are the orthonormal eigenvectors $\vec{w}_i$ of the covariance matrix of $\matrix{D}$ with the largest eigenvalues $\lambda_i$. The results of the PCA of the AIM configurations are displayed in Figs.~\ref{fig:expvar} and \ref{fig:unsupervised}. Figure~\ref{fig:expvar} shows the explained variance ratio \mbox{$\bar{\lambda}_i = \lambda_i /\sum_{j=1}^{L^2} \lambda_j$} of the first 25 principal components, where the $\lambda_i$ are sorted in descending order. 
The seven-dimensional subspace of the $L^2$-dimensional configuration space spanned by $\vec{w}_{1-7}$ explains more than $99.9\%$ of the variance in $\matrix{D}$. The first principal component is given by \mbox{$\vec{w}_1 = \tfrac{1}{L}[1,\dotsc,1]_{L\times L}$}, and it corresponds to the total magnetization. The next few leading components ($\vec{w}_2$--$\vec{w}_7$) appear in pairs with equal $\bar{\lambda}$ and are periodic along the $x$-direction with a period of $L$, $L/2$, and $L/3$ respectively. The pairwise occurrence of these components is required to describe the band structures in the `L+G' phase in the translationally invariant system. The smaller fluctuations in the local magnetization along the $y$-direction are represented by the higher prinicipal components $\vec{w}_{i>7}$. Each AIM configuration (i.e.,\ row of $\matrix{D}$) can be described by a set of projection coefficients $p_i$, which are the components of the AIM configuration in the lower-dimensional space spanned by the reduced set of principal \mbox{components (see Figs.~\ref{fig:unsupervised}(a)--\ref{fig:unsupervised}(c))}.\\

We denote $\left<p_i\right>$ as the fixed-temperature average of $p_i$. The temperature-dependence of $\left<p_i\right>$ allows one to separate the AIM configurations into three phases \mbox{(Figs.~\ref{fig:unsupervised}(d) and \ref{fig:unsupervised}(e))}. These phase boundaries are compared to those obtained by evaluating the mean magnetization per particle $\overline{m}$ and the liquid fraction 
\begin{equation}\label{eq:phi} 
\phi = \frac{1}{m_l L^2}\sum_i m_i \; ,
\end{equation} where $m_l$ is the magnetization of the liquid band, for the configurations in $\matrix{D}$. From the previous discussion, it is clear that $\left<|p_1|\right>/(\globaldensity L)$ is equal to the order parameter $\overline{m}$. Similarly, as $\left<\sqrt{p_2^2 + p_3^2}\right>$ is an indicator for the presence of large-scale inhomogeneities in the magnetization, it is nonvanishing for temperatures corresponding to the `L+G' phase. The maximum of $\left<\sqrt{p_2^2 + p_3^2}\right>$ allows us to infer the temperature for which the spatial liquid-gas ratio is equal to $1/2$, as the components with a period of $L$ dominate for that temperature.\\

Though the subspaces identified by PCA are readily understood, the separation into the different phases from its output turns out to be less straightforward. Additionally, the success of PCA in identifying the phases in the active Ising model does not imply that it leads to good results for other models. The reason for this is that, unlike nonlinear learning methods, it does not preserve local distances when projecting from a high-dimensional to a low-dimensional space. Figure~\ref{fig:unsupervised}(f) shows the result of dimensionality reduction applied to the data set $\matrix{D}$ of AIM configurations with a state-of-the-art nonlinear technique known as UMAP~\cite{mcinnes2018umap}. The UMAP algorithm assumes a manifold on which the original high-dimensional data are uniformly distributed, and it uses local fuzzy simplicial set representations to construct a topological representation of the data. It then searches for an optimal low-dimensional representation that has a fuzzy topological representation as similar as possible to the high-dimensional one. The algorithm is explained in more detail in Appendix~\ref{sec:UMAP}. This dimensionality reduction is implemented with the UMAP software package~\cite{mcinnes2018umap-software}. UMAP is highly efficient in uncovering the different phases of the AIM. Indeed, as becomes clear from Fig.~\ref{fig:unsupervised}(f), in the constructed representation with two UMAP components, UMAP-1 and UMAP-2, the AIM configurations clearly cluster in five well separated groups with specific temperature ranges. One group contains configurations with control parameter combinations belonging to the `G' phase. As is the case for PCA, the symmetry breaking in the `L' and `L+G' phases is uncovered by the UMAP algorithm, since it divides the configurations with positive and negative magnetization into a pair of clusters with the same temperature range. Remark that in Fig.~\ref{fig:unsupervised}(f) for both the `L' or `L+G' phases, the relative position of the two clusters with respect to the `G' phase is equal. Unlike PCA, UMAP is able to efficiently learn the translational invariance of the bands in the `L+G' phase and hence requires only two variables to classify the AIM configurations into the three phases. By identifying the temperature ranges of the different clusters in the UMAP subspace, we can now easily infer the transition points between the three phases.

\section{Supervised learning} \subsection{Classification} The presented analysis clearly showed that unsupervised learning can determine the temperature boundaries for the different AIM phases at a fixed global density $\globaldensity$. In this section, we demonstrate that supervised learning trained with phase-labeled AIM configurations at a fixed $\globaldensity$ is capable of predicting the phase boundaries in a wide range of $\globaldensity$ values that it did not encounter during the training procedure. To this end, a convolutional neural network (CNN) is first trained on a data set of AIM configurations generated at $\globaldensity = 3$ (see Appendix~\ref{sec:appCNN} for network architecture and training details). This training is supervised, since the configurations are now labeled with their respective phase, using the results of the unsupervised UMAP approach.
Given an input configuration $I$, with a magnetization and density `channel', the network assigns a class score $S^I_c$ to each of the three phases \mbox{$c=$ `L', `G', or `L+G'}, from which the probability $P^I_c$ for the configuration to belong to phase $c$ can be found after a softmax operation:
\begin{equation}\label{eq:softmax} P^I_c = e^{S^I_c}/\sum_{c'} e^{S^I_{c'}}.\end{equation} During training, the loss function $\mathcal{L}$ is minimized by optimizing the model's weights $w \in \mathcal{W}$ which connect the different layers. For each AIM configuration $I$, the loss function reads
$\mathcal{L}_I(Q^I;\mathcal{W}) = \mathcal{H}(Q^I, P^I(\mathcal{W}))$,
where 
\begin{equation}
\mathcal{H}(Q^I,P^I(\mathcal{W})) = -\sum_c Q^I_c\log(P_c^I(\mathcal{W}))
\end{equation}
is the cross entropy between the predicted ($P^I(\mathcal{W})$) and the true ($Q^I$) class probabilities. An additional \mbox{L2-regularization} function with strength $\lambda$ is included in the total loss function: \begin{equation}\label{eq:loss} \mathcal{L} = \tfrac{1}{N}\sum_I \mathcal{L}_I + \lambda \sum_{w \in \mathcal{W}} w^2. \end{equation} The addition of the regularizing term reduces the magnitude of the weights $\mathcal{W}$. As a result the learned features tend to be more simple for $\lambda > 0$, as it prevents the model from focusing its decision boundaries on highly specific features of the individual training examples. Instead, the addition of regularization forces the model to find the more general features of the underlying data distribution \cite{goodfellow2016deep}.

\begin{figure}[b]
\includegraphics{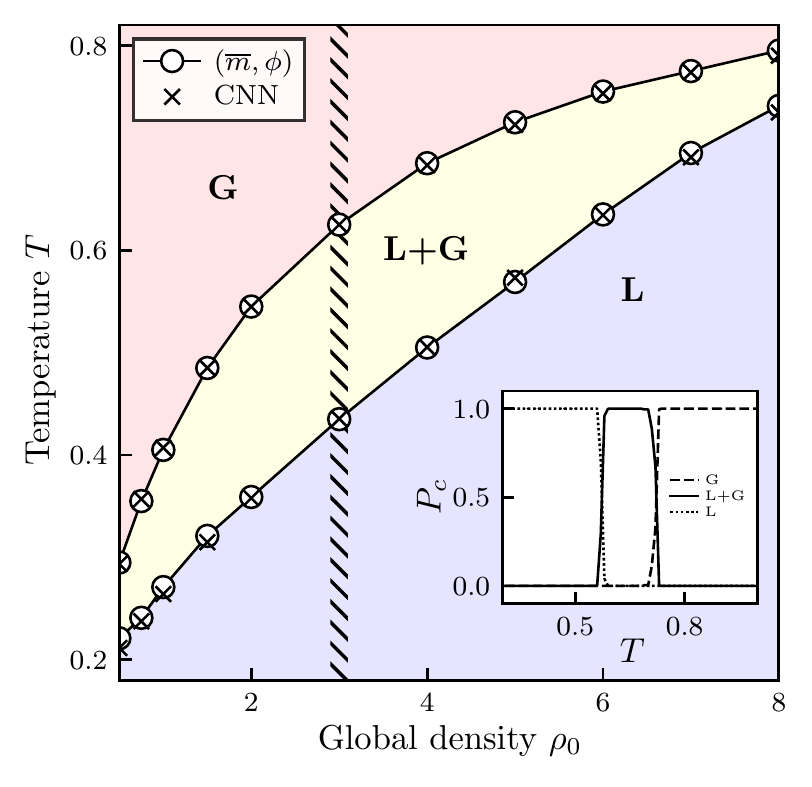}
\caption{The phase boundaries  of the AIM for an $81 \times 81$ lattice in the $(\globaldensity, T)$-plane. 
The circles are the phase boundaries obtained by evaluating the mean magnetization per particle $\overline{m}$ (Eq.~(\ref{eq:m})) and the liquid fraction $\phi$ (Eq.~(\ref{eq:phi})) for a grid of $(\globaldensity, T)$ values. The crosses are inferred with a CNN that is only trained on configurations with $\globaldensity = 3$ (hatched region). Inset: The CNN's prediction of the temperature dependence of the average probability $P_c$ (Eq.~(\ref{eq:softmax})) of an AIM configuration to belong to the `L', `G', `L+G' phases for $\globaldensity = 5$.}
  \label{fig:boundaries}
\end{figure}%

\subsection{Interpretability}
In order to figure out the features that the CNN has captured, we first feed the model with AIM configurations of unknown phase labeling sampled at $\globaldensity$ values not included during training. Networks failing to predict the phase boundaries under those circumstances are likely to have learned trivial features from the $\globaldensity=3$ data, {\it e.g.} the local magnetizations $m_i$ crossing a threshold.  The phase boundary between two phases $c^\prime$ and $c^{\prime\prime}$ is inferred from the temperature for which the predicted class probabilities $P_{c^\prime}$ and $P_{c^{\prime\prime}}$ coincide~\cite{Carrasquilla2017, vanNieuwenburg2017}. Networks trained without regularization ($\lambda = 0$) can perfectly classify unseen AIM configurations sampled at the same control parameters ($\globaldensity =3$ and $T \in [0.2,1.0]$) used during the training phase of the CNN. Yet, they often fail in assigning the proper phase for configurations with combinations of $(\globaldensity, T)$ that were not included during training. In such cases, the minimum of the loss function focuses heavily on details specific to the $\globaldensity=3$ configurations, and the neural network has failed to grasp the more general features of the AIM. The addition of a small regularizing term to the loss function in Eq.~(\ref{eq:loss}) limits the model's complexity, but it does not impact its classification accuracy on the $\globaldensity = 3$ data set. Remarkably, as shown in Fig.~\ref{fig:boundaries}, networks trained with $\lambda > 0$ are able to extrapolate the boundaries they have learned at $\globaldensity = 3$ to a range of densities $ 0.5 \le \globaldensity \le 8.0$ extending over more than one order of magnitude. Hence, this gives a first hint that for $\lambda > 0$, the CNN extracts the more physically relevant characteristics, as it gains the potential to accurately determine phase boundaries at $\globaldensity$ far away from the training set. On top of that, its inferred extrapolation of the phase boundaries is robust, meaning that the results depend little on the initial weight parameters and choices with regard to the training set.\\

In other words, each run of the optimization routine with a different regularization strength results in a particular set of weight parameters. Although these neural networks reach a similar classification accuracy on configurations with control parameter combinations they have encountered during training, only a selected subset succeeds in making a physically relevant classification. The latter illustrates the pitfalls of merely using classification accuracy for model selection, without scrutinizing the learned features. Indeed, networks performing well for configurations with control parameters included in training may fail to capture the physics required for a proper classification of unseen $(\globaldensity, T)$-combinations. \\
\begin{figure}[b]
\includegraphics{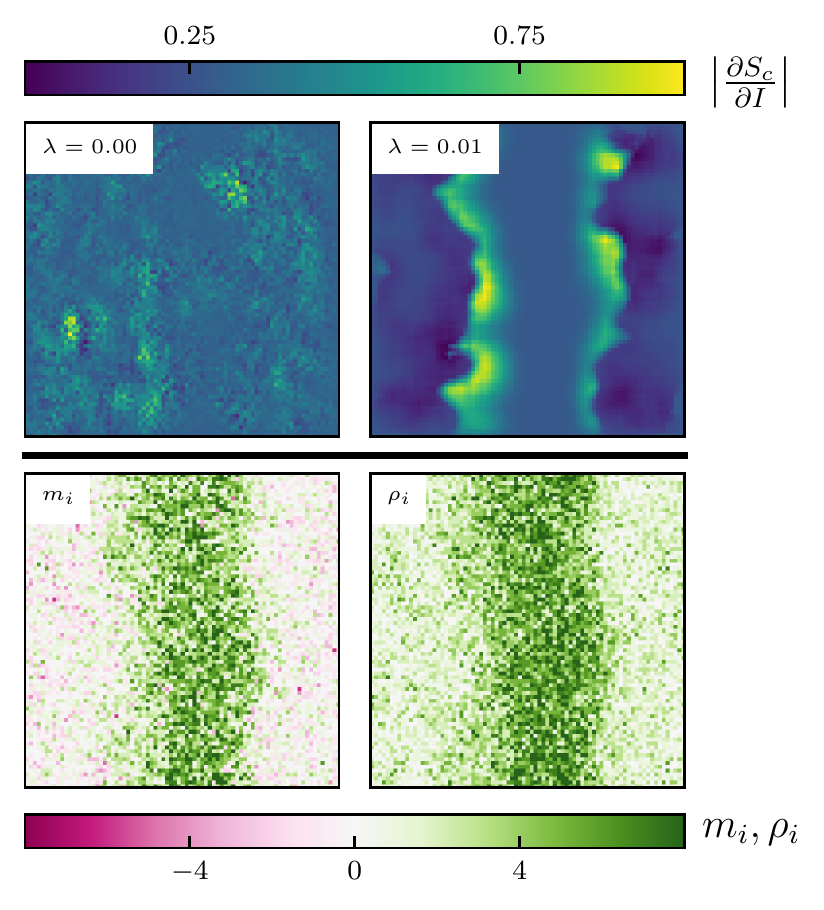}
\caption{
Saliency maps and physical properties of an $81 \times 81$ AIM configuration in the `L+G' phase with $\globaldensity = 3$, \mbox{$T = 0.56$}. Top panels:  $|\partial S_c/\partial I|$, normalized between 0 and 1, for a network trained with L2-regularization strength $\lambda = 0$ and $\lambda = 0.01$. Bottom panels: local magnetization and local density.}
  \label{fig:saliency}
\end{figure} 

We now address the issue of whether the neural network bases its decision on the phase classification on physically relevant features, and the specific role the hyperparameters play in this. For this purpose, we turn towards interpretability methods developed to gain insight into ``black-box'' classifiers in the context of image classification. One readily available tool is a saliency map~\cite{simonyan2013deep}, which identifies the pixels on which a classifier builds its decision. Given an AIM configuration $I$ of phase $c$, to which the network assigns a class score $S_c$, we compute the quantity $\left|\partial S_c / \partial I\right|$ through back propagation, where we take the maximum value of this gradient over the local magnetization and density input channels for each lattice site (see also Fig.~\ref{fig:cnn}). As a result, we can highlight the regions of $I$ that heavily impact the classification. Those regions are interpreted by the CNN as phase-characteristic and---if captured correctly by the neural network---should be reminiscent of the physical features. To illustrate the potential of saliency maps in phase characterization, we first train a network on AIM configurations for all global densities shown in Fig.~\ref{fig:boundaries}. For the `L' and `G' phases, the gradient $|\partial S_c/\partial I|$ attains only small values, which reflects that the model has captured the homogeneity of these phases. The `L+G' phase is much more challenging with regard to phase classification. A prototypical saliency map for the  `L+G' phase is shown in Fig.~\ref{fig:saliency} for a vanishing and nonvanishing regularization strength $\lambda$. The magnitude of $\lambda$ strongly impacts whether the algorithm identifies physical features. Without any regularization ($\lambda = 0$), the network's decision is clearly built on very local characteristics. Once regularization is turned on, it succeeds in identifying global emergent properties---for the AIM these are the diffuse edges between liquid and gas. As illustrated in Fig.~\ref{fig:filters}, this can also be observed in the filters of the first convolutional layer. These saliency maps hence give a clear insight into why the addition of a regularization term to the loss function of Eq.~(\ref{eq:loss}) is necessary for a proper extrapolation of the phase boundaries in Fig.~\ref{fig:boundaries}.
\\

\section{Conclusion} We have demonstrated that a sequential application of unsupervised and supervised machine learning is a powerful instrument to infer and characterize the phase diagram of a liquid-gas transition, without any a priori knowledge of its phase boundaries. Advanced dimensionality reduction methods, such as UMAP, clearly cluster system configurations into the physical phases and recognize the presence of symmetry breaking. By feeding a convolutional neural network with phase-labeled configurations, we demonstrated that well-designed neural networks, trained to learn the phase boundaries for fixed control parameters, are capable of extrapolating the phase boundaries to complete the phase diagram for a wide range of control parameters. Thereby, it is of crucial importance to properly select the network architectures and hyperparameters. Indeed, we have demonstrated that neural networks with a comparable classification performance can either learn physically relevant features or meaningless properties. The addition of a regularizing term to the loss function is an instrument to discriminate between these networks. By employing interpretability tools, such as saliency maps, the strength of the regularization can be connected to the locality of the features captured by the neural network. Here, we have illustrated that after adding saliency maps to the deep learning procedure, we can extract the core physical features, e.g.,~the phase-characteristic magnetization and density patterns.\\

\begin{acknowledgments}
We are indebted to Benjamin Vandermarliere, Andres Belaza, Ken Bastiaensen and Wesley De Neve for useful discussions. The computational resources (Stevin Supercomputer Infrastructure) and services used in this work were provided by the VSC (Flemish Supercomputer Center).
This work was supported by Ghent University, Research Foundation Flanders (FWO-Flanders) and the Flemish Government -- department EWI. T.~Vieijra is supported as an `FWO-aspirant' under contract number FWO18/ASP/279.
\end{acknowledgments}
\FloatBarrier
\revappendix
\section{UMAP}\label{sec:UMAP}

The UMAP algorithm \cite{mcinnes2018umap, mcinnes2018umap-software} contains two major steps: constructing a (fuzzy) topological representation for the  high-dimensional data, followed by optimizing a low-dimensional representation of it.\\

The process of finding a cover for a manifold on which the data lies gets facilitated if the data are uniformly distributed on the manifold. Hence, given the data, one can first define a Riemannian metric that accomplishes this requirement. In practice, the algorithm considers each data point with its nearest neighbors, and computes a metric locally by normalizing the volume of the ball that includes these data points---from now on referred to as the metric spaces. In this way, every data point is assigned its own independent distance measure, valid in its vicinity. Accordingly, the local metric spaces are incompatible and need to be merged in order to form a consistent global structure. UMAP solves this issue by translating each of the metric spaces to a fuzzy simplicial set and then taking the fuzzy union over the family of these sets. In practice, the algorithm constructs 1-simplices (edges) connecting the data points. Each 1-simplex is assigned a weight $w \in [0,1]$, which can be intuitively understood as the probability that an edge exists between two points. By following these steps, a fuzzy topological structure for the high-dimensional data is formed. In essence, the result of this approach is a neighborhood graph.\\

The construction of a neighborhood graph for the data in the low-dimensional space follows a similar pattern. In comparison to the high-dimensional space, it is simpler because one knows the distance metric and the manifold $\mathbb{R}^n$, with $n$ the chosen number of dimensions for the projection. The fuzzy topological structures of the low-dimensional and high-dimensional representations should be as similar as possible. Given the set of possible \mbox{1-simplices} and their assigned weights $\mathcal{W}_h$ and $\mathcal{W}_l$ in the high- and low-dimensional representation respectively, the measure for this similarity is the cross entropy  $\mathcal{H}(\mathcal{W}_h, \mathcal{W}_l)$. With the aid of stochastic gradient descent, the algorithm updates the weights $\mathcal{W}_l$ to minimize $\mathcal{H}(\mathcal{W}_h, \mathcal{W}_l)$.

\section{CNN architecture and training}\label{sec:appCNN}
\begin{figure}[b]
\centering
\includegraphics[width=\columnwidth]{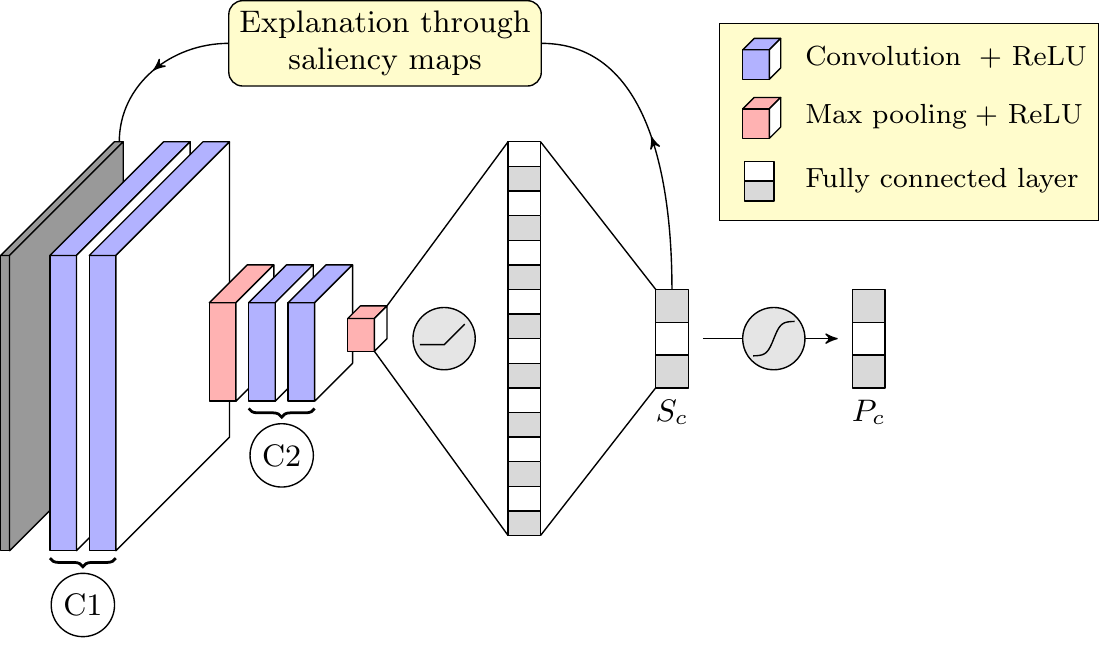}
	\caption{Architecture of the CNN used for inferring the phase boundaries in the control parameter space of the AIM.}
	\label{fig:cnn}
\end{figure}

The convolutional neural network architecture used to determine and characterize the phase boundaries is shown in Fig.~\ref{fig:cnn}. The input layer consists of two channels: magnetization and density. The first two convolutional layers (C1) each have 6 filters with a $(5 \times 5)$ kernel and have a ReLU activation function. These layers are followed by a max pooling layer, with a $(3 \times 3)$ kernel and stride 3. Pooling is included to reduce the model complexity and to make the observed features less orientation- and scale-dependent. The next two convolutional layers (C2) also contain 6 filters with ReLU activations, but now with a $(3 \times 3)$ kernel, and they are followed by the same max pooling operation. The flattened feature vector is then sent through a fully-connected network, where the first layer has 16 hidden nodes with ReLU activations. The output layer has three nodes, one for each of the three different phases, and a softmax activation. The network is defined by a total of $\left(\tfrac{32}{27}L^2+1939\right)$ weights and biases, which are trained using an Adam optimizer with learning rate $\alpha = 10^{-3}$. We found that the learned classification was rather insensitive to the value of the learning rate $\alpha$. The data is split into a training set~($60\%$ of the total data), a validation set~(20$\%$), and a test set~(20$\%$). To avoid overfitting on the training set, the loss function is evaluated on the validation set after every training epoch. The model with the lowest loss on the validation set is kept. When no decrease in validation loss is detected for $100$ consecutive training epochs, the training is terminated (``early stopping'') and the network is evaluated on the independent test set. The neural network and its training are implemented using TensorFlow~\cite{tensorflow2015-whitepaper} and Keras~\cite{chollet2015keras}.\\

\begin{figure}[h]
\centering
\includegraphics[width=\columnwidth]{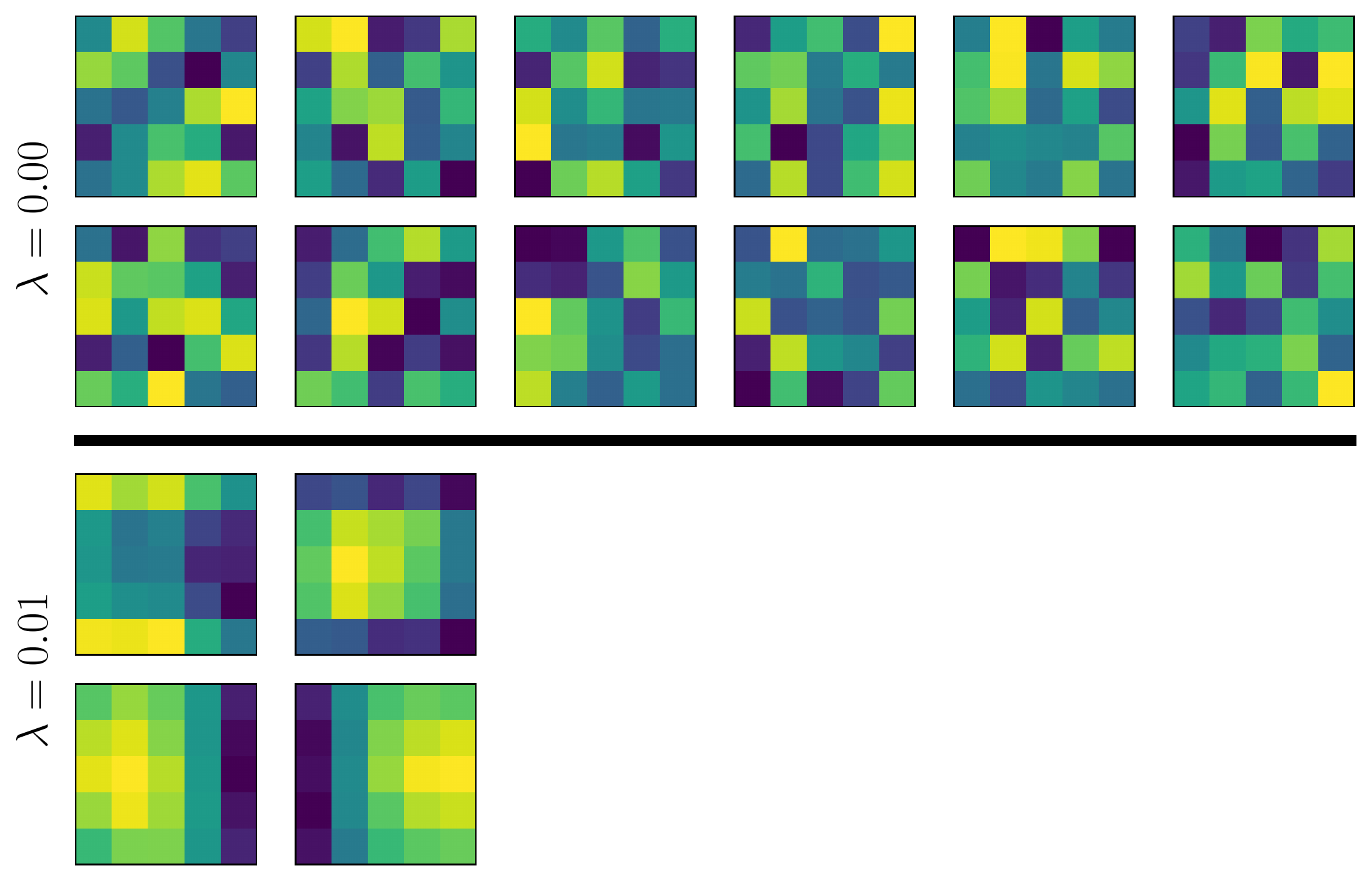}
\caption{Normalized kernels of the first layer of the CNN after training. Only kernels with absolute values of the weights larger than $0.01$ are shown. The top (bottom) two rows correspond to $\lambda=0$ ($\lambda=0.01$). The top and bottom row for each $\lambda$ represent the filters operating on the magnetization and density channel, respectively. \label{fig:filters} }
\end{figure}

In addition to the saliency maps, we interpret the inner workings of the CNN by visualizing the kernels of the first convolutional layer in Figure~\ref{fig:filters}. This clearly illustrates that the first layer of the CNN trained with $\lambda=0.01$ detects  more robust features (e.g.,~gradients in the local density values) compared with the nonregularized network.\\
\newpage

\FloatBarrier
\bibliography{bibliography.bib}
\end{document}